\documentclass[12pt,a4paper,]{article}
\usepackage{graphicx,psfrag}
\usepackage{multimedia}
\usepackage{color}
\usepackage{bm}
\usepackage{comment}
\usepackage[utf8]{inputenc}
\usepackage[T1]{fontenc}

\newcommand{\beqn}{\begin{eqnarray}} 
\newcommand{\eeqn}{\end{eqnarray}} 
\newcommand{\no}{\nonumber}
\newcommand{\halb}{\frac{1}{2}}

\newcommand{\one}{{\mathbf 1} }
\newcommand{\aaa}{\frac{\sqrt{3}}{6} }
\definecolor{darkred}{rgb}{0.8, 0., 0.}

\begin{document}

\title{A family of energy stable, skew-symmetric finite difference schemes on collocated grids}
\author{Julius Reiss}
\maketitle

\begin{abstract}
A simple scheme for incompressible, constant density flows is presented, which
avoids odd-even decoupling for the Laplacian on a collocated grids. Energy stability is implied by maintaining strict
energy conservation. Momentum is conserved. Arbitrary order in space and time can easily be 
obtained.  The conservation properties hold on transformed grids.  
    

\end{abstract}

\section{Introduction}
\label{intro}
The odd-even decoupling is one of the central issues when simulating incompressible flows. 
It refers to the fact that the Laplace operator calculated from a discrete gradient and  a discrete divergence operator 
decomposes into two Laplace operators living on super-grids, when central derivative operators are used.  
If, for example (in one dimension), the gradient at grid position $i$ is defined by $(G p)_i = (p_{i+1} -p_{i-1})/(2\Delta x)$ and 
the divergence by $ (D u)_i = (u_{i+1} -u_{i-1})/(2\Delta x)$, the implied  Laplace operator is $(D G p)_i = (p_{i-2}-2 p_{i}+p_{i+2} )/(2\Delta h)^2$. 
Thus, it connects only every second point on the grid. This decoupling leads to severe convergence problems. 

The most simple cure are upwind schemes, where the derivative stencil is chosen asymmetric in accordance with the flow direction. 
This leads to high numerical dissipation especially when using low order derivatives. 
If this is not acceptable, two strategies to circumvent this problem are commonly followed. Either the usage of a staggered grids or the increasingly popular  
the Rhie-Chow interpolation \cite{RhieChow1983}. In the latter a small regularization term is added, which suppresses the decoupling. See \cite{Wesseling2001} for a discussion of 
the different methods.  

Here a different approach is presented. It is based on the observation that asymmetric spatial derivatives do not necessarily imply 
that the resulting scheme is violating the energy conservation. The discretization is obtained by using the skew-symmetric form 
of the transport term \cite{Morinishi1998,VerstappenVeldman2003}. The resulting approach is elegant due to its simplicity.   
The scheme presented here is constructed  analogous to a scheme for  {\it compressible} flows \cite{ReissSesterhenn2014}. 
Since in compressible flows an equation of motion for the pressure can be formulated, odd-even decoupling is 
a minor issue.  

An alternative scheme which similarly builds on the skew-symmetric form, 
but instead uses a special interpolation procedure to avoid decoupling on collocated grids was recently proposed by \cite{TriasLehmkuhlOlivaPerz-SagarraVerstappen2014}.

A first version of the following scheme was presented in \cite{Reiss2012}. Here additionally high order time integration schemes,
conservation on arbitrary grids and simulations on transformed grids are included.

\subsection{Skew-symmetric schemes}  

Skew-symmetric schemes are schemes which conserve the kinetic energy by design. To illustrate the basic idea, consider an
 equation of motion for the vector quantity $u$ 
$$
\partial_t u = { A} u, 
$$
where $A$ is a skew-symmetric matrix, meaning  $A^T=-A$. To  change of the norm or the kinetic energy can be derived as     
\beqn
    u^T \partial_t  u 
&=& \halb \partial_t u^T u  \no\\ 
&=& -  u^T A u  = 0 .
\eeqn
The last steps follows from  $ z = u^T A u =  (u^T A^T u ) = - u^T A u = -z $, so that $z = 0 $. In other words quadratic forms of skew-symmetric matrices 
always vanish, as the terms $A_{i,j}u_iu_j$ and $A_{j,i}u_ju_i = - A_{i,j}u_ju_i$ cancel pairwise.

To utilize this concept for the numerical evaluation of the Navier-Stokes equation, the momentum equation is discretized in such a way that the nonlinear transport 
term is skew-symmetric, directly implying the conservation of the kinetic energy by this term. 
Numerical damping and the artificial change of kinetic energy are basically different perspectives on the same phenomenon. Thus a conserved kinetic energy 
can be seen as zero numerical damping, or as an energy stable scheme. The physical damping of course reduces the kinetic energy. In  reality  it is converted  to heat or internal energy, 
which is usually not accounted for in the description of incompressible flows, because the change in temperature is typically very small.

In the following it will be shown that a skew-symmetric discretization can easily be  constructed even with asymmetrical, i.e. not skew-symmetric  derivatives.  Thus
it is possible to construct schemes without numerical damping with asymmetrical matrices. The asymmetrical form in turn allows to avoid the odd-even decoupling on collocated grids.   
The  discussion  in this work is restricted to the periodic case without boundaries.  
   
\section{Derivation of the Scheme}
\label{sec:scheme}
Starting point is the standard Navier-Stokes or momentum equation with constant density $\varrho \equiv 1$  
\beqn
\partial_t u_\alpha + \partial_{x_\beta} u_\beta u_\alpha +  \partial_{x_\alpha} p  &=& \nu \Delta u_\alpha.
\label{e:nsDiv}
\eeqn
Here, $u_\alpha$ are the compontens of the velocity field, $p$  is the pressure and $\nu$ the kinematic viscosity. $\Delta$ denotes the Laplace operator.
Greek letters mark  the summation over space directions $\alpha,\beta = 1,2,3$. Summing convention is assumed. 
The Navier-Stokes equation is complemented by the continuity equation or solenoidal condition  
\beqn
\partial_{x_\alpha} {u_\alpha}	=0, 
\label{solenoidal} 
\eeqn
to describe incompressible flow. To satisfy this algebraic condition the pressure has to be determined accordingly.
 The defining pressure Poisson equation is obtained by applying the solenoidal condition (\ref{solenoidal}) on the
 momentum equation (\ref{e:nsDiv}), see below.        
 
 The Navier-Stokes  equation  can be rewritten with the help of the  continuity equation as
\beqn
\partial_t u_\alpha +    u_\beta 
\partial_{x_\beta}  u_\alpha +  \partial_{x_\alpha} p  &=& \nu \Delta u_\alpha. 
\label{e:nsCon}
\eeqn
Summing equations  (\ref{e:nsDiv}) and (\ref{e:nsCon}) leads  to the skew-symmetric form of 
the Navier Stokes equation: 
\beqn
\partial_t u_\alpha + \halb { \left( \partial_{x_\beta} u_\beta\cdot +  u_\beta 
\partial_{x_\beta} \cdot \right)} u_\alpha +  \partial_{x_\alpha} p  &=& \nu \Delta u_\alpha \label{e:nsSkew}
\eeqn
The derivatives in in the parenthesis are understood to act on the velocity to the right of it, which is denoted by a dot. 
 
The {\it discretization} can be done by replacing the functions $u_\alpha(x,y,z)$ and $p(x,y,z)$ by discrete values at 
$(x_i,y_j,z_k) = (i\cdot\Delta x, j\cdot\Delta y , k \cdot\Delta z) $, where equidistant spacing is assumed for now; and further by replacing the derivatives
 in (\ref{solenoidal}) and (\ref{e:nsSkew}) by derivative matrices.  
Later two different derivatives have to be utilized. In anticipation of these findings we insert the 
different derivatives in accordance with the use of the gradient ($G_\alpha$) or divergence($D_\alpha$). 
The discrete equations are       
\beqn
{D_\alpha} {u_\alpha}	&=& 0, \label{e:solenoidalDisc} \\
\partial_t u_\alpha + {\halb \left( D_\beta U_\beta  +  U_\beta G_\beta  \right)}
 u_\alpha +  G_\alpha p  &=& \nu L u_\alpha.\label{e:nsSkewDisc}
\eeqn
The derivatives are assumed to have stencils only in the discretized direction, as usual. They can be written as standard one dimensional matrices with 
  the help of the Kronecker product. In two dimensions the divergence would be 
$D_x \equiv D_1 = {\mathbf I}^{2}   \otimes d_1  $ and  $D_y \equiv D_2 = d_2 \otimes {\mathbf I}^{1} $. Here,  $d_\alpha$ is a one dimensional discretization of the
divergence and ${\mathbf I}^{N_\alpha}$ is the unity matrix in the given direction. All fields are sorted accordingly in one dimensional vectors. 
Capital letters mark pointwise multiplication, which can be represented by a diagonal matrix with the corresponding field on the diagonal.   
The abbreviation for the transport term 
\beqn
  {D^{\bf u}} = {\halb \left( D_\beta U_\beta  +  U_\beta G_\beta  \right)} \label{e:Du}
\eeqn
will be used. 
The symmetry of the transport term is found to be  
\beqn
 {D^{\bf u }}^T   =  \halb 
	(U_\beta D_\beta^T     +  G_\beta^T U_\beta  ) \equiv -   D^{\bf u }, 
\eeqn
which is skew-symmetric, provided that 
\beqn
 D_\alpha ^T    = -    G_\alpha \label{e:DTG}  
\eeqn
It is  further assumed that the stencil is the same for every grid point.  
The kinetic energy is conserved; multiplying (\ref{e:nsSkewDisc}) by $u_\alpha^T$ gives  
\beqn
 u_\alpha^T \partial_t u_\alpha + {u_\alpha^T  D^{\bf u}  u_\alpha} + {u_\alpha^T G_\alpha p} &=& \nu u_\alpha^T L u_\alpha. 
\eeqn
The skew-symmetric transport term is zero by construction ${u_\alpha^T  D^{\bf u}  u_\alpha} = -{u_\alpha^T  D^{\bf u}  u_\alpha}   = 0 $. The 
pressure work is 
\beqn
u_\alpha^T G_\alpha p =  - p^T   D_\alpha  u_\alpha =0.
\eeqn
It vanishes for incompressible flows due to the solenoidal condition. Thus, the change of kinetic energy 
\beqn
\halb \partial_t u_\alpha^T u_\alpha  = \nu u_\alpha^T L u_\alpha 
\eeqn
is given by the physical friction alone. This term is usually discretized in a symmetric fashion and negative semi-definite, so that it can only reduce the kinetic energy.
 
The  {pressure Poisson equation} is derived by applying the divergence on (\ref{e:nsSkewDisc}), yielding with the help of (\ref{e:solenoidalDisc}) 
\beqn
D_\alpha G_\alpha p  &=&  D_\alpha\left(- {D^{\bf u}}
 u_\alpha +   \nu L u_\alpha\right) 
\eeqn
We now come to the central point of this work. The restriction on the derivatives  (\ref{e:DTG}) is sufficient to conserve energy. 
This restriction does not imply that any of the  derivatives has to be central. 
A decoupling of implied Laplace operator $  \Delta = D_\alpha G_\alpha $ can simply be  avoided by using asymmetric matrixes. 
 As an example the left side derivative for the divergence $ (D  u)_i  = (u_{i}  - u_{i-1})/\Delta x  $, and $ (D  u)_i  = (u_{i}  - u_{i-1})/\Delta x $   for the gradient 
is valid and leads obviously to the standard second derivative for the Laplacian $(DG p)_i = (p_{i-1}-2p_{i} +p_{i+1}) /(\Delta x)^2 $.  
Higher order derivatives can be used, as long as they are  asymmetric. This discretization should be clearly distinguished from an upwind scheme. 
First, as mentioned in the introduction, upwind stencils are changing with the flow direction, which is not the case in the present scheme.   
Secondly the asymmetrical stencil does not imply numerical damping as derived before.
While both terms in the transport (\ref{e:Du})  separately do change the kinetic energy, the contributions exactly cancel in combination.

The {\it conservation of momentum } is less obvious,  as  (\ref{e:nsSkewDisc}) is not in divergence form. It is checked by summing over (\ref{e:nsSkewDisc}), which is the discrete analog of the integral. 
The sum is  represented by a vector   $\one$ where all components  are one:  
\beqn
\partial_t \one^T u_\alpha + {\halb \one^T \left( D_\beta U_\beta  +  U_\beta G_\beta  \right)}
 u_\alpha + \one^T  G_\alpha p  &=& \nu  \one^T L u_\alpha. 
\eeqn   
Since the stencil is assumed to be the same for every grid point,  we trivially have the  telescoping sum property, i.e. the sum over the columns of the derivatives vanish. Thus 
$\one^T D_\alpha = \one^T G_\alpha   = \one^T L = 0 $ .  The only remaining term  is 
\beqn
\one^T   U_\beta G_\beta    u_\alpha  =   u_\beta G_\beta    u_\alpha =  u_\alpha^T G_\beta^T u_\beta ,  
\eeqn
which reduces with (\ref{e:DTG}) to the  solenoidal equation (\ref{e:solenoidalDisc}) and is therefore zero. Momentum is thus conserved  
\beqn
\partial_t \one^T u_\alpha = 0 \no. 
\eeqn    
\section{Transformed Grids}

The scheme can be extended to curvilinear grids.  This is done in the same manner as in \cite{ReissSesterhenn2014,ReissSesterhenn2011}. 
The main question is, if momentum and energy conservation strictly hold, since both build here on the solenoidal condition, while in 
the compressible case an extra energy equation is available. The conservation demands, that divergence of the metric terms vanishes.   
A proper definition of the metric terms especially in three dimensions leads to strict conservation of both terms. 
 
The transformation is given as $x_\alpha = x_\alpha(\xi_1,\xi_2,\xi_3 )$. Following \cite{ThompsonWarsiMastin1985}, one can  define the local base as 
\beqn 
 {\bf e}_\alpha= \partial_{\xi_\alpha} 
\left(\begin{array}{c} x \\ y\\ z\end{array}  \right),  \label{e:nabla} 
 \eeqn
and obtain two  equivalent forms of the Nabla operator  
\beqn
\nabla {\varphi } &=& \frac 1 {J} \sum_{\alpha} \partial_{\xi_\alpha} {\bf m}_\alpha {\varphi }  
=
\frac 1 {J} \sum_{\alpha}  {\bf m}_\alpha  \partial_{\xi_\alpha} {\varphi }. \label{nabNonCons}
\eeqn
The Jacobian is $J=({\bf e}_1\times {\bf e}_2)\cdot {\bf e}_3$. Vector valued metric factors are introduced, which are calculated as 
\beqn
{\bf m}_\alpha  = {({\bf e}_\beta\times {\bf e}_\gamma)} \qquad \alpha,\beta,\gamma\mathrm{\quad cyclic.} 
\eeqn
Since the components of (\ref{e:nabla}) contain the derivatives in physical space, we can use these 
two forms to discretize the gradient as 
\beqn
 D_\beta  =   \bar D_\alpha M_{\alpha,\beta}  
 \eeqn
and 
\beqn
  G_\beta  =   M_{\alpha ,\beta} \bar G_\alpha  ,  
  \eeqn
where $\bar D_\beta$ and $\bar G_\beta $  are operators in the computational or $\xi$-space. As the
discretization is chosen to be equidistant, the same derivatives as used previously in the Cartesian case 
 can be used in the computational space.  

The discrete, transformed  Navier-Stokes  equation is 
\beqn
\frac 1 J  \bar D_\beta M_{\alpha,\beta }{u_\alpha}	&=& 0, \\
J \partial_t u_\alpha + 
\halb \left(  \bar D_\gamma M_{\beta,\gamma}  U_\beta  +
  U_\beta  M_{\beta,\gamma }   \bar 	G_\gamma  \right)
 u_\alpha +   M_{\alpha,\gamma} \bar G_\gamma	 p  &=& \nu \tilde L u_\alpha. \label{e:ns_trans} 
\eeqn
One can simplify the equation by defining effective convective velocities 
\beqn
\tilde u_\beta = M_{\alpha,\beta }{u_\alpha}
\eeqn
to obtain 
\beqn
 \bar D_\beta M_{\alpha,\beta }{u_\alpha}	&=& 0 \\
J \partial_t u_\alpha + 
{D^{\bf \tilde u}}
 u_\alpha +   
{\bar G_{x_\alpha}} 	 p  &=& \nu \tilde L u_\alpha. 
\eeqn
The factor $J$ was dropped in the continuity equation. 
The transport term has the same structure as on the Cartesian grid.  
\beqn
D^{\bf \tilde u} = 
  \halb \left(  \bar D_\gamma   \tilde U_\gamma  +
               \tilde  U_\gamma \bar G_\gamma  \right) . 
\eeqn
and the pressure gradient is $ \bar G_{x_\alpha} = M_{\alpha,\gamma} \bar G_\gamma$.  
The friction term can be discretized in the same way as the Laplacian of the pressure equation, see below. 

The derivatives in computational space satisfy, as before, in physical space,  
\beqn
\bar D_\beta = - \bar G_\beta^T ,   
\eeqn
which is again the key for the energy conservation. 

The pressure Poisson equation becomes 
$$
{ \bar L } \; p  =  { \bar D_\beta \frac{M_{\alpha,\beta}   M_{\alpha ,\gamma}}{J} \bar 
G_\gamma} 
  \; p  =  
   \bar D_\beta M_{\alpha,\beta } \left[\left( - D^{\bf \tilde u} u_\alpha  +  \nu \tilde L u_\alpha \right)/J \right].   
$$
A factor of $J$ is canceled on both sides.  A discrete approximation of the Laplacian is thus $L = 1/J \bar L $. 
The friction term in the source term should not be omitted, as could be done in the continuous case. 
Due to the  metric factors the divergence operator does only approximately commute with the Laplacian. 

In two dimensions, which will be considered in the numerical examples, we have 
\beqn
\tilde u_1 = \tilde u = \phantom{-} u y_\eta - v  x_\eta \\
\tilde u_2 = \tilde v = -u y_\xi  + v x_\xi
\eeqn
and the pressure gradient is
\beqn
\bar G_x p  = \left(\phantom{-} y_\eta \bar G_1 -  y_\xi \bar G_2 \right) p \\
\bar G_y p  = \left(-x_\eta \bar G_1 +  x_\xi \bar G_2  \right) p.   
\eeqn

In the following we derive the conservation properties. 
For the {\it energy conservation} we have to multiply (\ref{e:ns_trans}) with $u_\alpha$. The transport term is 
still skew-symmetric, thus it drops out as before. Assuming a vanishing physical friction $\nu = 0 $,  we arrive at  
\beqn
\partial_t \left( u_\alpha^T J  u_\alpha\right)/2 &=&
       - u_\alpha^T { M_{\alpha,\gamma} \bar G_\gamma	 p } \no\\
 &=& - p^T   \bar G_\gamma^T	M_{\alpha,\gamma} u_\alpha  \no\\
 &= &  p^T   { \bar D_\gamma	M_{\alpha,\gamma} u_\alpha}  =  0  
\eeqn
by the help of continuity equation. A non-zero friction would create the extra term 
\beqn
\nu u_\alpha^T\tilde L u_\alpha &=& 
\nu u_\alpha^T  \bar D_\beta \frac{M_{\alpha,\beta}   M_{\alpha ,\gamma}}{J} \bar  G_\gamma  u_\alpha\no\\
&=& -  \left(M_{\alpha,\beta} \bar D_\beta  u_\alpha\right)^T \frac 1 J \left(M_{\alpha ,\gamma} \bar  G_\gamma  u_\alpha \right)    
\eeqn
which is due to its symmetry obviously negative semi-definit, if we  assume the same structure for the friction Laplacian  as in the the pressure Poisson equation. 
 
The {\it momentum conservation} is checked by summing over (\ref{e:ns_trans}), using the telescoping sum property 
 \beqn
\partial_t \left(  1^T J u_\alpha\right)  &=&  -\halb   
  u_\beta^T  M_{\beta,\gamma }   \bar 	G_\gamma  u_\alpha -   {\bf 1}^T { M_{\alpha,\gamma} \bar G_\gamma p}  \\
&=&  + \halb u_\alpha   \bar 	D_\gamma M_{\beta,\gamma }   u_\beta +  p^T   \bar G_\gamma^T	m_{\alpha,\gamma} \no\\
  &=& - p^T    { \bar  D_\gamma	m_{\alpha,\gamma}}  .
 \eeqn
Thus, we have to evaluate the divergence  of the metric factors. \\
In {\it two dimensions} the metric factors are 
$ {\mathbf m}_1 = (y_\eta,-x_\eta, 0 ) $ and ${\mathbf m}_2 = (-y_\xi,x_\xi, 0 )^T $ 
\beqn
 \bar D_\gamma	m_{\alpha,\gamma}=
 	\left(  \begin{array}{c} \bar D_\xi y_\eta - \bar D_\eta y_\xi \\  -  \bar D_\xi x_\eta + \bar D_\eta x_\xi \end{array}\right) , 
 \eeqn
which is zero if the metric factors are calculated with the derivative of the divergence:   
\beqn
 y_\xi   =
    \bar D_\xi y\qquad
 y_\eta = 
    \bar D_\eta y.
\eeqn 
If the metric factors would be calculated by the gradient energy conservation would be only approximate.\\
For the general {\it three dimensional} case we obtain  
\beqn
    &&\bar D_\gamma^T	m_{\alpha,\gamma}    \no
    =	\\
    &&\left(  \begin{array}{c} 
    \bar D_\xi ( y_\eta z_\zeta - y_\zeta z_\eta ) 
    + \bar D_\eta (  y_\zeta z_\xi -y_\xi z_\zeta ) 
    + \bar D_\zeta ( y_\xi z_\eta -  y_\eta  z_\xi  ) 
    \\ \dots \end{array}\right) 
  \eeqn
In the analytical case this can be shown to be zero by using the product rule on all terms. 
As the product rule does not hold in the discrete, this form would break the stric conservation of momentum. 
This problem is easily circumvented by the nice trick of Thomas and Lombard \cite{ThomasLombard1979}:  
The metric factors above can analytically be rewritten as 
\beqn
y_\eta z_\zeta - y_\zeta z_\eta &=& ( y_\eta z)_\zeta - (y_\zeta z)_\eta  \no \\
y_\zeta z_\xi -y_\xi z_\zeta    &=& (y_\zeta z)_\xi   - (y_\xi   z)_\zeta \no \\
y_\xi z_\eta -  y_\eta  z_\xi   &=& (y_\xi   z)_\eta -  (y_\eta  z)_\xi \no 
\eeqn 
If the discretization builds on this form the divergence of the metric vanishes for any discretization. 
This rewriting also improves the quality of the simulation strongly, as reported in \cite{VisbalGaitonde2002}.

\section{Time Stepping}
\label{time}
To keep the strict conservation the time discretization has to respect the momentum and energy conservation. 
The implicit midpoint rule is an adequate, second order  choice \cite{VerstappenVeldman2003}. This implicit midpoint rule belongs
 to the class of Gauss collocation Runge-Kutta methods, which all conserve quadratic invariants. The general theory can be found in 
\cite{HairerLubichWanner2006}. The fourth order rule is defined by the Butcher table \\
\begin{center}
\begin{tabular}{c|cc}
$ \halb -\aaa $&$\frac 1 4$ & $\frac 1 4 -\aaa$ \\
$ \halb +\aaa $&$\frac 1 4 +\aaa$ & $\frac 1 4$ \\
\hline
               &$\halb$&$\halb$
\end{tabular}.
 \end{center}
In \cite{BrouwerReissSesterhenn2012,BrouwerReissSesterhenn2014}  it is shown how to use this schemes for the skew-symmetric, compressible Euler equations. 
For the incompressible case these schemes can be used without adaptation. This was already reported by \cite{Sanderse2013}.
Also higher order Gauss collocation Runge-Kutta methods can be used, but might be of little practical interest.
The pressure has to be determined for each implicit Runge-Kutta step. Splitting schemes do destroy the strict conservation. 
However, splitting gives in our experience well working an stable schemes.   
For the sake of completeness we sketch the the prove of energy conservation. The time stepping is 
\beqn
u_\alpha^{n+1}
 &=& u_\alpha^{n} +\frac{\Delta t}{2} (k^-_\alpha + k^+_\alpha) \label{e:unp} \\
u_\alpha^-& =&   u_\alpha^{n} + \Delta t\left(  \frac 1 4    k^-_\alpha + \left( \frac 1 4 - a \right ) k^+_\alpha \right) \label{e:um} \\
u_\alpha^+& =&   u_\alpha^{n} + \Delta t\left( \left( \frac 1 4 + a \right ) k^-_\alpha +  \frac 1 4   k^+_\alpha \right)  \label{e:up} 
\eeqn 
where  $ k^\pm_\alpha =\mathrm{rhs}(u^\pm) $ is the right hand side of the momentum equation, i.e. all terms but the time derivative and $a = \aaa$. The kinetic energy at $t^{n+1}$ is  
calculated with (\ref{e:unp})
\beqn
(u^{n+1}_\alpha)^T u^{n+1}_\alpha &=& \no 
(u^{n}_\alpha)^T u^{n}_\alpha  \\&&+ \frac{\Delta t}{2} (u^{n}_\alpha)^T (k^-_\alpha + k^+_\alpha)  +    \frac{(\Delta t)^2}{4} (k^-_\alpha + k^+_\alpha)^T(k^-_\alpha + k^+_\alpha) 
\eeqn
expressing $u_\alpha^{n}$ in the term linear in $\Delta t$ second by (\ref{e:um}) in front of $k^-_\alpha$ and by (\ref{e:up}) in front of $ k^+_\alpha$ all quadratic terms cancel:
\beqn
(u^{n+1}_\alpha)^T u^{n+1}_\alpha =  
(u^{n}_\alpha)^T u^{n}_\alpha + \frac{\Delta t}{2} ( (u^{-}_\alpha)^T k^-_\alpha + (u^{+}_\alpha)^Tk^+_\alpha).
\eeqn 
The term linear in $\Delta t$ vanishes as shown in the previous section, since the combinations are evaluated at the same time. 
The momentum conservation is right away obtained from summing (\ref{e:unp}). 
Of course, the pressure has to be determined for each implicit Runge-Kutta step.

\section{Numerical Example}
\label{Numerical}
We present two cases to prove the principle soundness of the approach. 
Both cases are two dimensional and periodic. 
The first one is a vortex pair on a transformed grid. 
The second case consists of three merging vortices on an Cartesian grid. 
Third  and fifth order stencils are used. 
The fifth order divergence  is given by the stencil 
$(0,  3 ,-30, -20, 60, -15,  2)/60$, the gradient is obtained by transposing. Time stepping algorithms of second and fourth order are used.   
The pressure equation can still be solved by a pivoted LU decomposition.     
The start solution is made divergence-free as usual by an initial Chorin projection step \cite{Chorin1968}.
The implicit time stepping is solved by a fix point iteration. This works  well for moderate time steps, as reported in \cite{BrouwerReissSesterhenn2014}.

\subsection{Vortex pair on a transformed grid}

\begin{figure}

\begin{center}
\includegraphics[width=0.45\linewidth, clip =true, viewport =60 0 490 420]{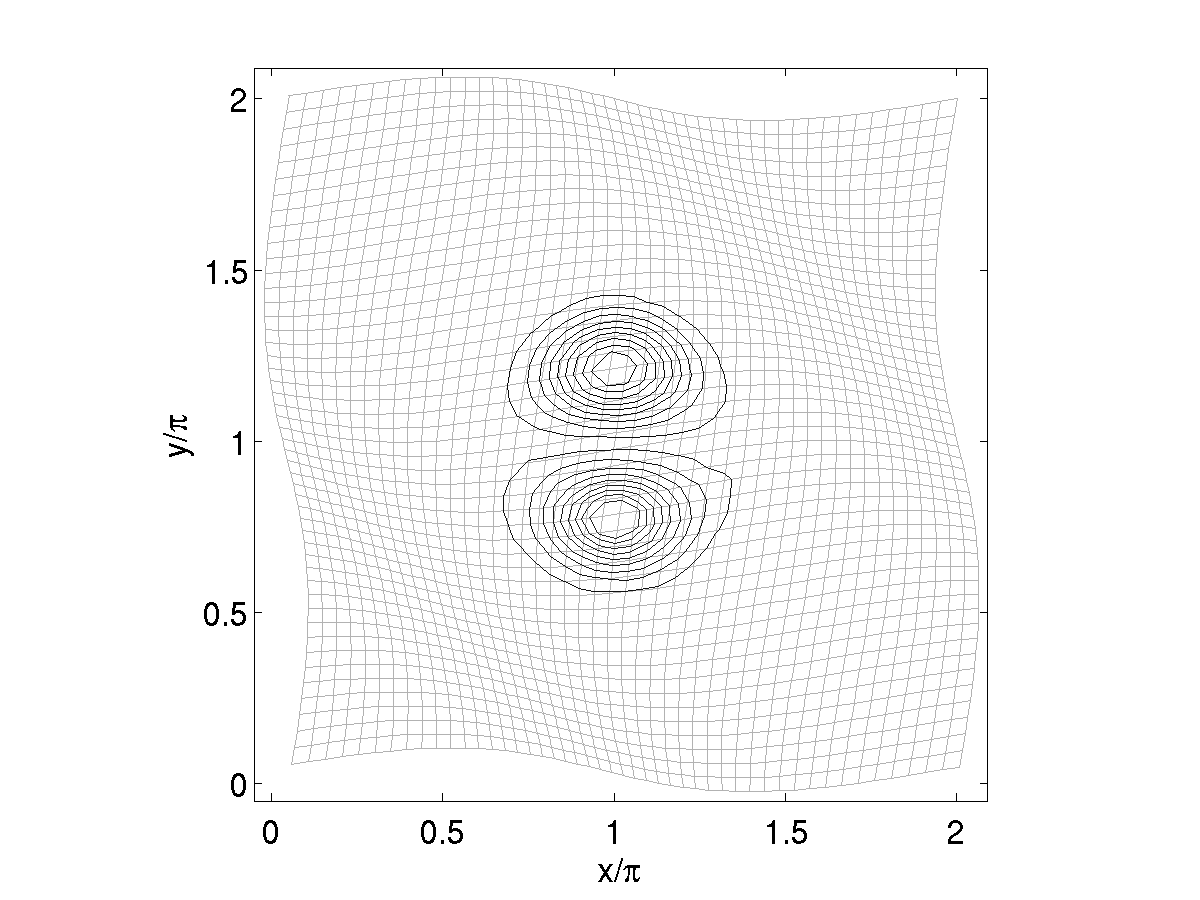}
\includegraphics[width=0.45\linewidth, clip =true, viewport =60 0 490 420]{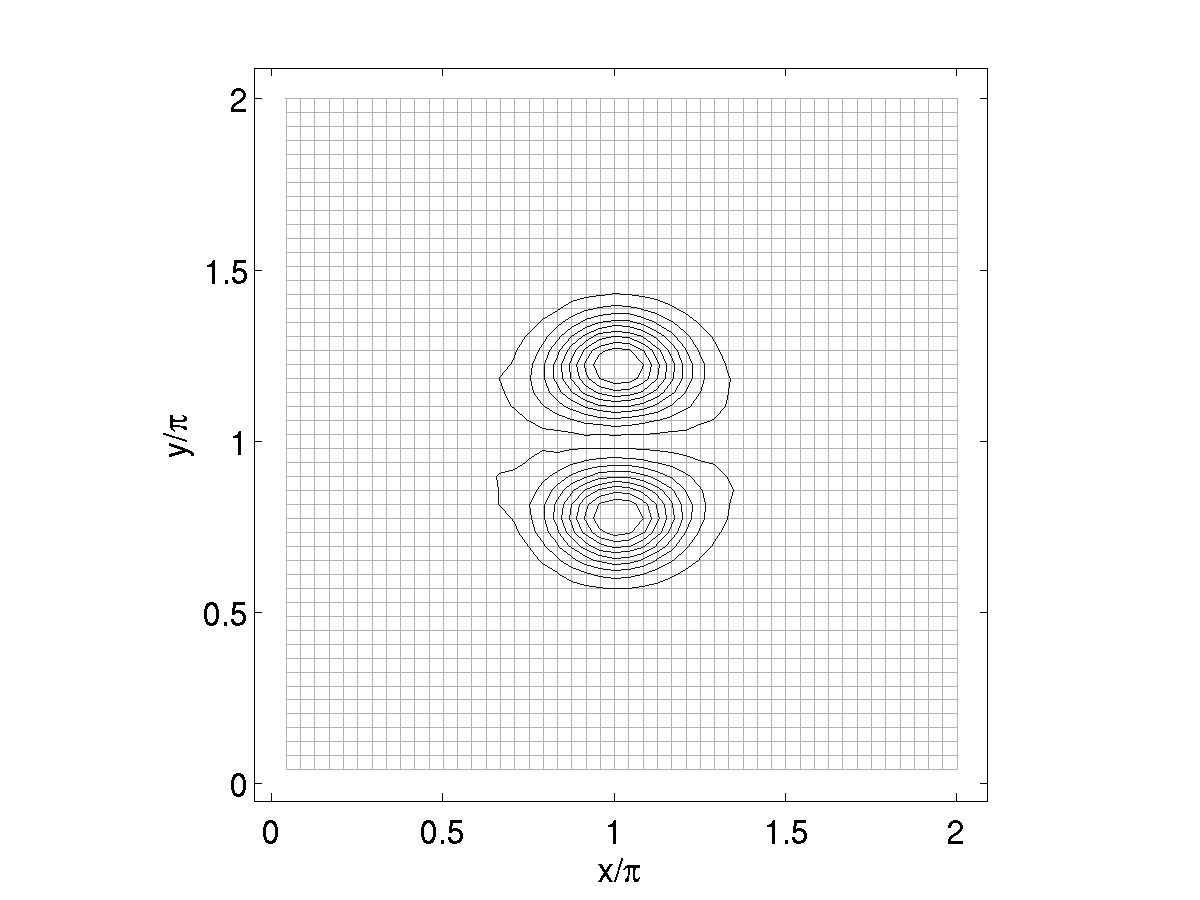}
\end{center}
\caption{The rotation of a vortex pair on a strongly non-orthogonal grid, level lines are separated by $0.4$, where most inner line of the top vortex is $-6.9$ and $6.9$ for the lower vortex. 
The vortex pair is shown after traveling through the full (periodic) domain. The vortex pair is only slightly influenced by the the strong distortion, due the fifth order derivatives. 
The quality of the most outer vortex lines is due to small gradients of $\omega$ strongly limited by the resolution. }
\label{fig:vortexPair}       
\end{figure}

\begin{figure}
\begin{center}
\includegraphics[width=0.45\linewidth, clip =true, viewport =30 0 550 500]{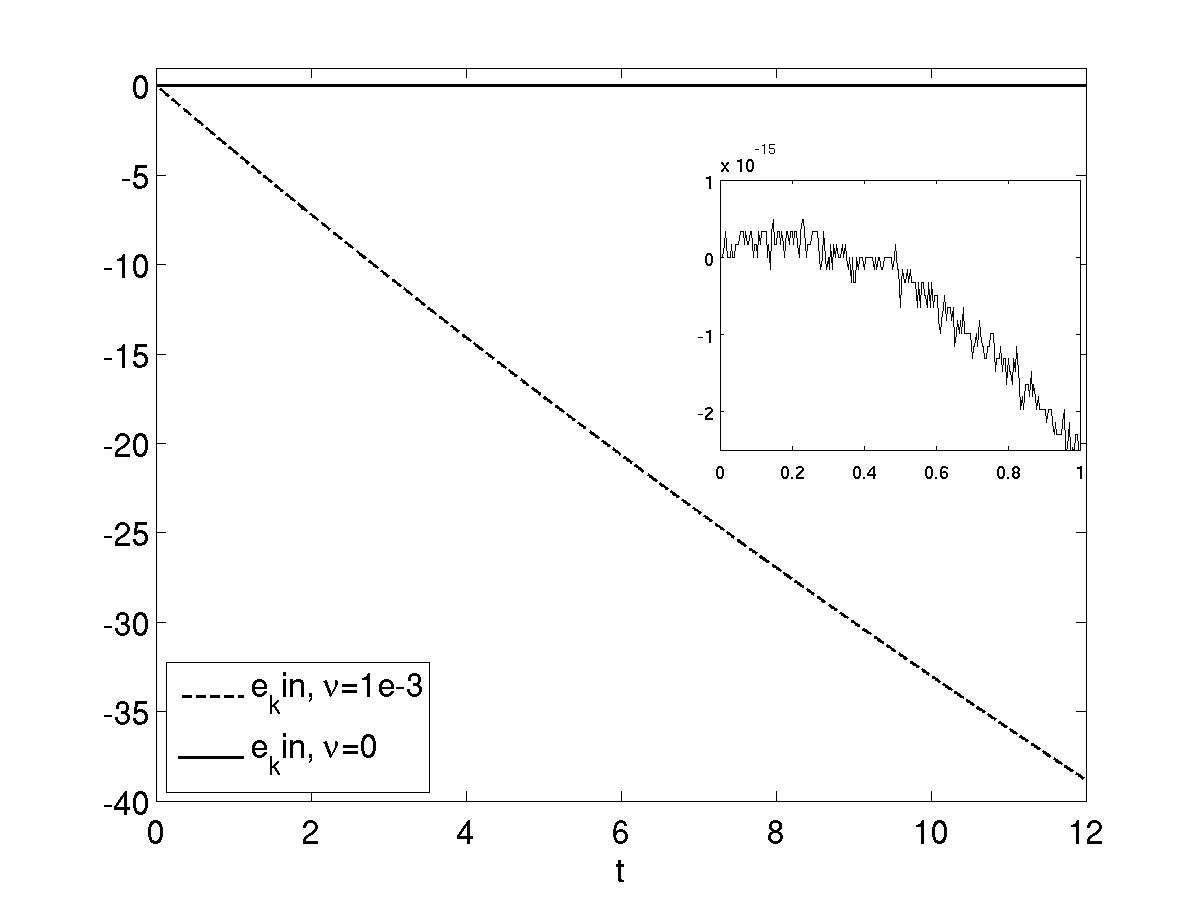}
\includegraphics[width=0.45\linewidth, clip =true, viewport =30 0 550 500]{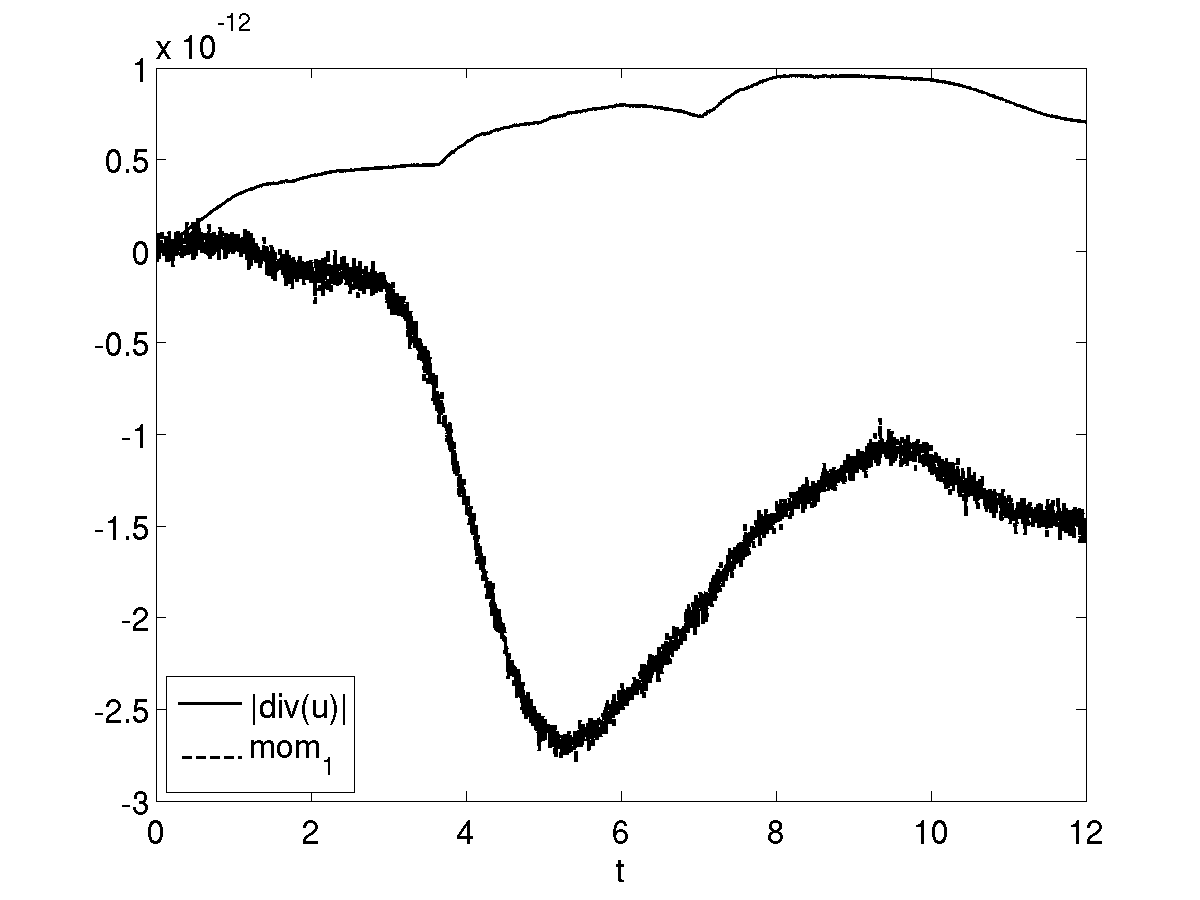}
\end{center}
\caption{Left: The kinetic energy is reduced only by physical friction. Strict conservation of kinetic energy if found for $\nu =0$. The relative change (inset) is of the order of $10^{-15}$. Right: The divergence is numerically zero,  the momentum is conserved.}
\label{fig:momEn}       
\end{figure}

The quality of the scheme on distorted grids is examined by the transport of a vortex pair on a strongly distorted grid. The vortex pair is given by 
\beqn
\omega = \alpha \left[ e^{-((x-x_1)^2+(y-y_1)^2)/\beta^2 }-e^{ -((x-x_1)^2+(y-y_1)^2)/\beta^2 } \right] 
\eeqn
with $x=(L/2,L/2)$, $y=(0.4\cdot L,0.6\cdot L)$.
The amplitude $\alpha=10$ and the width  $\beta=L/14$.
The reference size and the size of the computational space is $L=2\pi$. The transformed grid is  given by $x = \xi + k \sin(\xi+\eta) $ and $y = \eta +  k \sin(\xi+\eta)$, with $k=0.2$. The discretization is $N_\xi, N_\eta =48,49$.      
The spatial discretization error is of fifth order. The time integration is the fourth order Gauss method, with $\Delta t =4\cdot 10^{-3}$. 

The vortices are allowed to travel once through the full domain, crossing the lines of strong distortion.
It is expected that higher order derivatives reduce grid effects, as for a perfect derivative the grid transformation is analytically exact.   
Due to the fifth order derivative we find a good behavior despite this distortion.  
The conservation properties are shown in fig. \ref{fig:momEn}. Change of momentum and energy are zero within the numerical accuracy zero for the frictionless case. 
Friction reduces the kinetic energy as expected.

\subsection{Three vortices test case}

\begin{figure}
\begin{center}
\includegraphics[width=0.45\linewidth, clip =true, viewport =60 0 490 420 ]{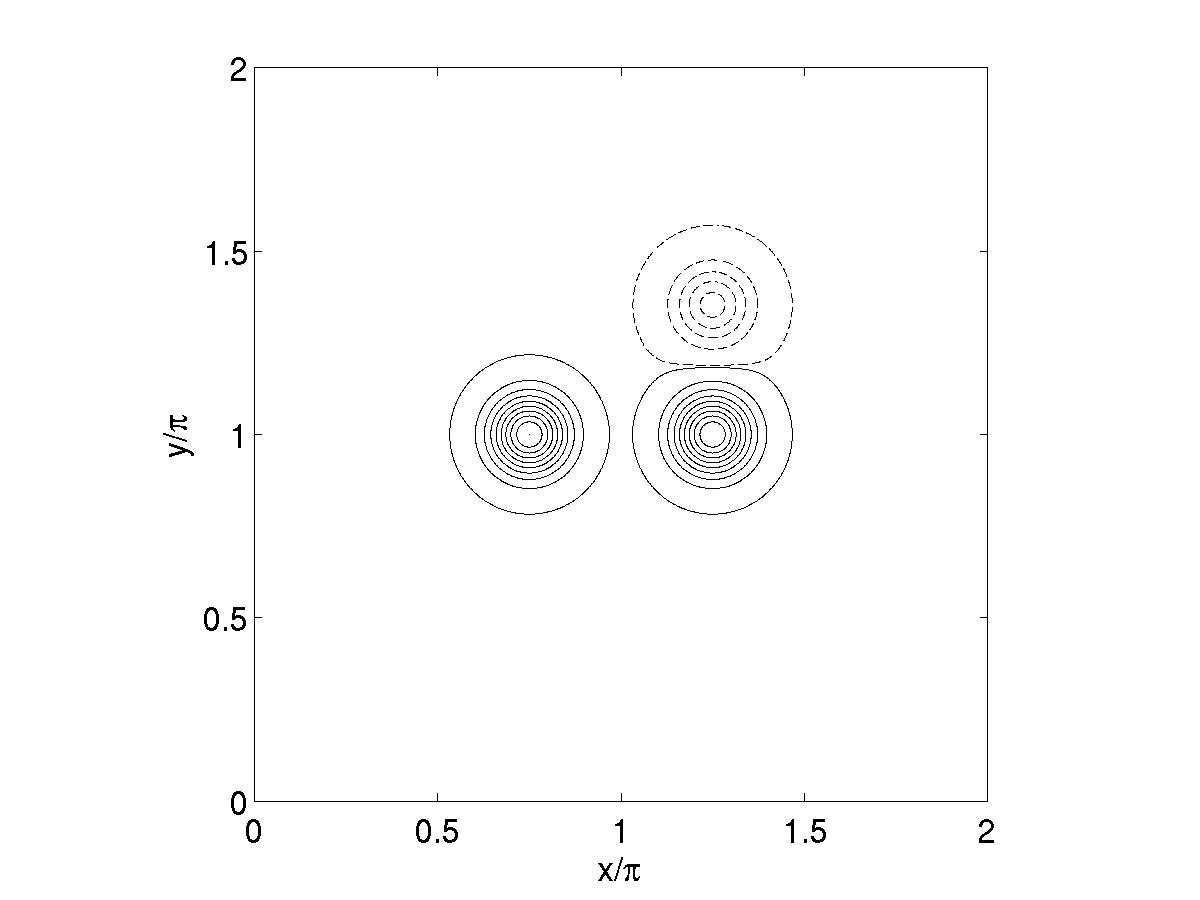}
\includegraphics[width=0.45\linewidth, clip =true, viewport =60 0 490 420]{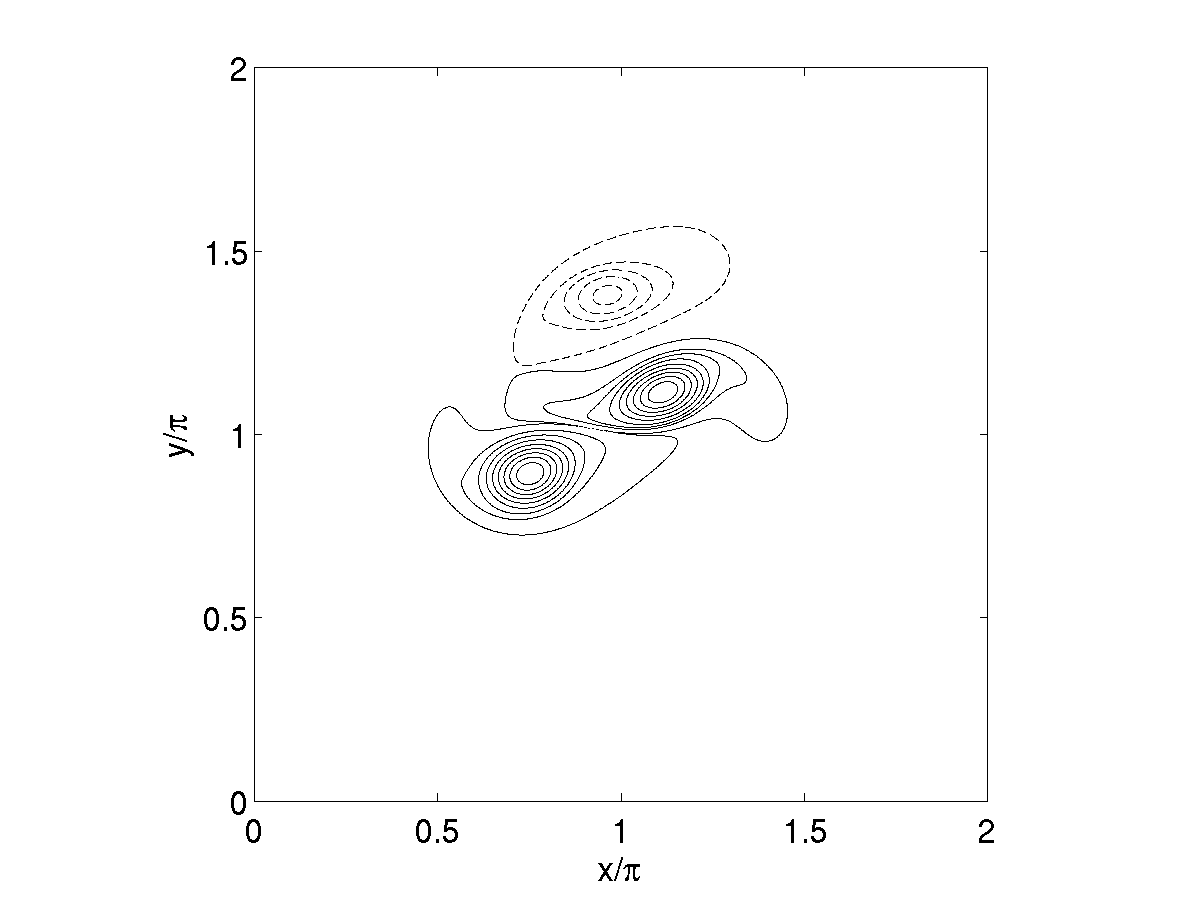}

\includegraphics[width=0.45\linewidth, clip =true, viewport =60 0 490 420]{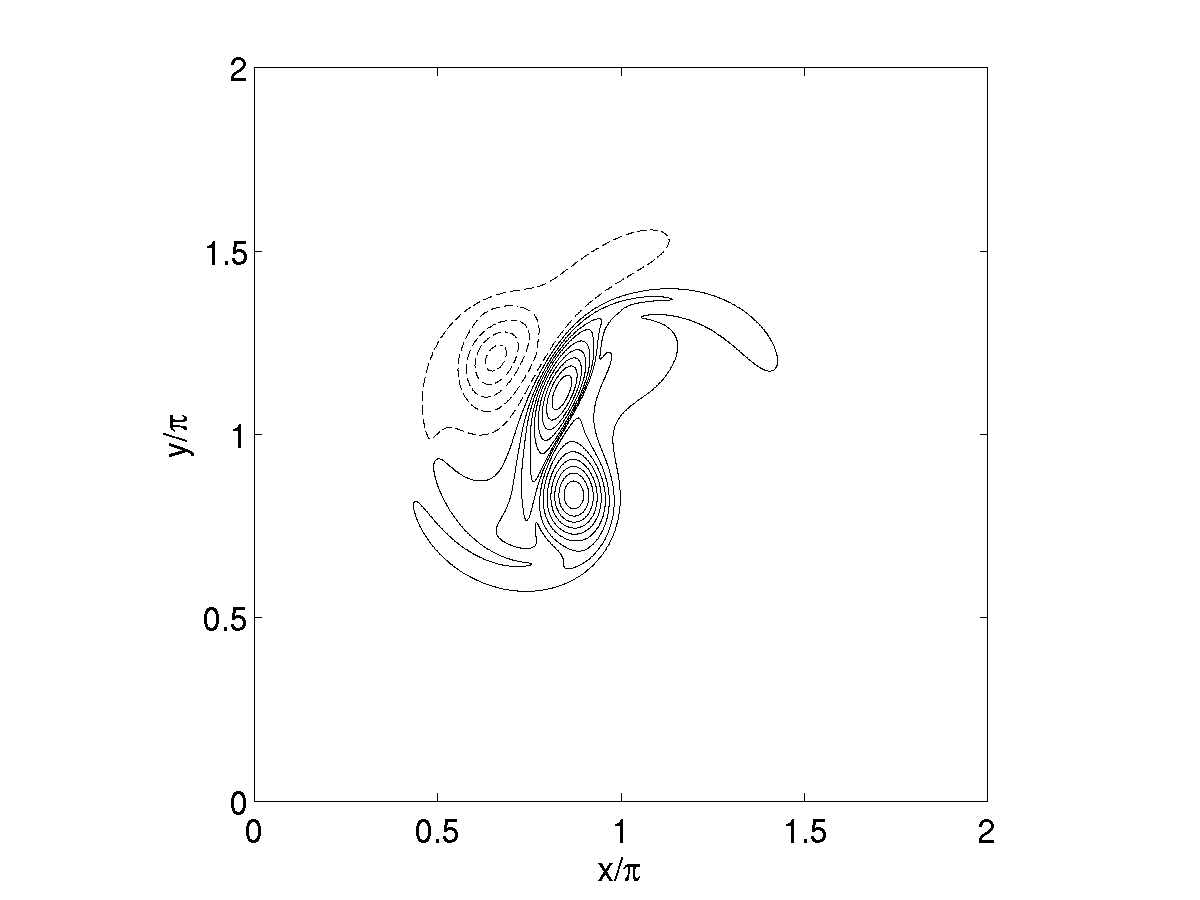}
\includegraphics[width=0.45\linewidth, clip =true, viewport =60 0 490 420]{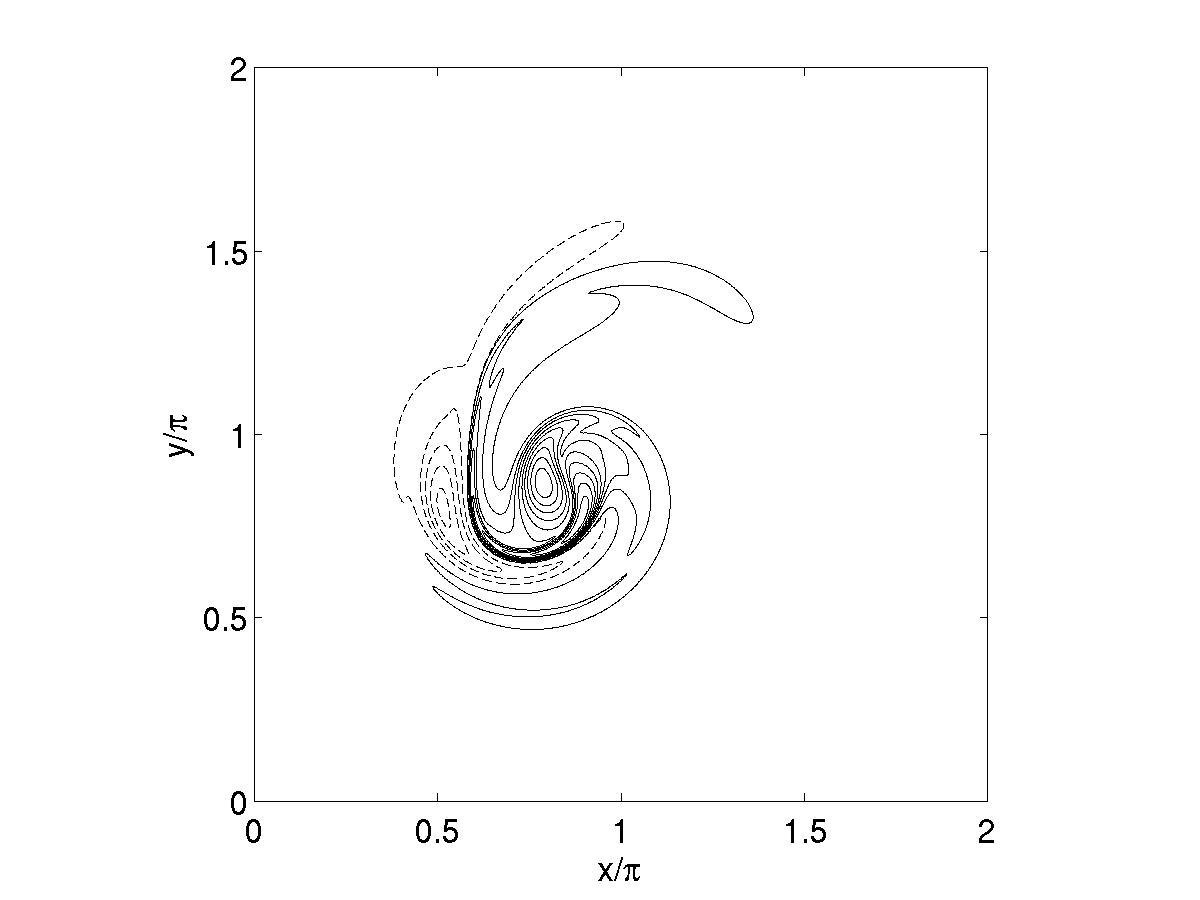}

\includegraphics[width=0.45\linewidth, clip =true, viewport =60 0 490 420]{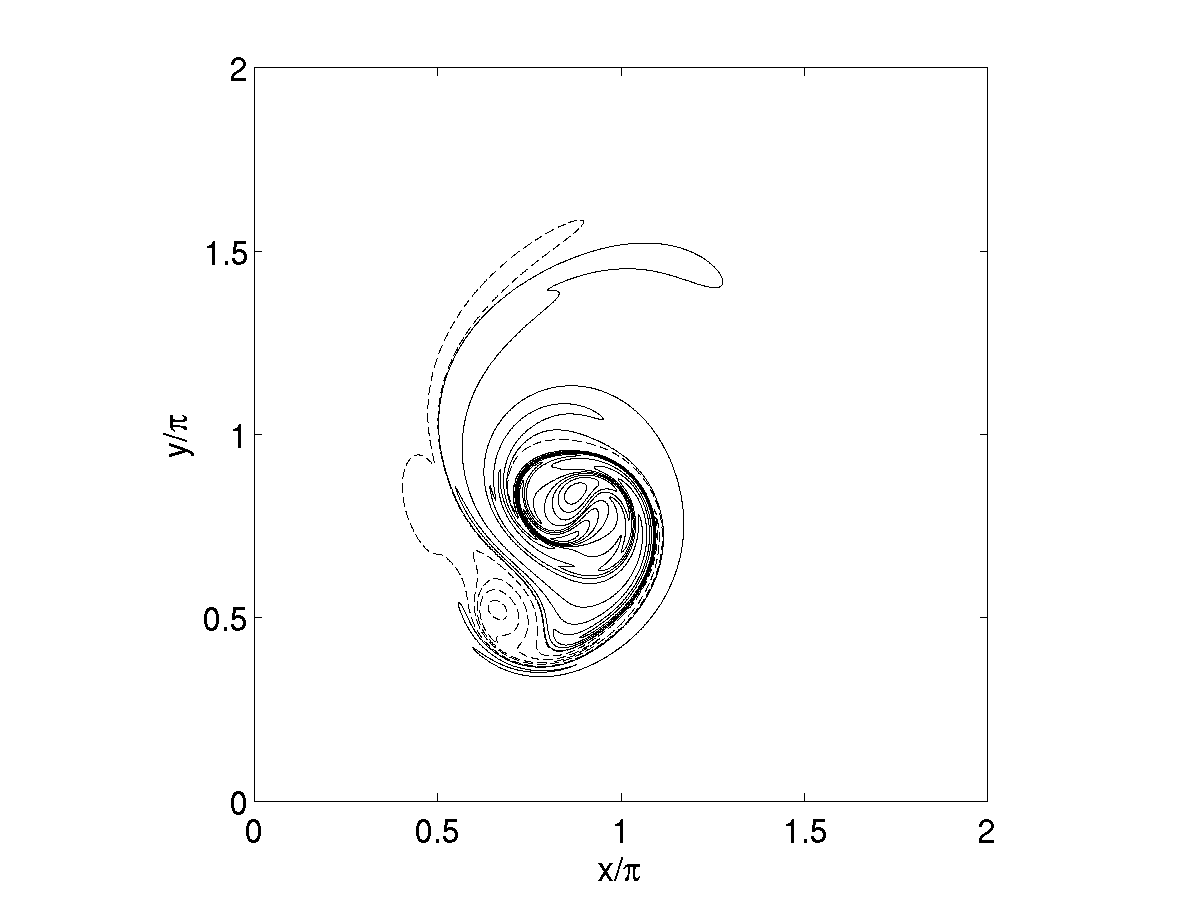}
\hspace{0.45\linewidth} 

\end{center}
\caption{Three merging vortices at times $t=0,5,10,15$ and $t=20$ starting from upper left, upper right. The plots shows $\mathrm{rot}(u)-\omega_s$, with $\omega_s=-0.038$. 
The negative values are dashed contour lines from $(-\pi,-\pi/100)$ and the positive values are solid lines from $(\pi/100,\pi)$. The maximal negative contour line is $-1.0577$. }
\label{fig:threeVortex}       
\end{figure}

In this test case two of three vortices are merging, see \cite{KevlahanFarge1997}. 
A periodic and quadratic domain with a size of $2\pi$ is used.    
The vortices are given by  $\omega = \mathrm{rot}(u) = \omega_0 + \sum_k \alpha \exp(r_k^2/\beta^2)$ with 
$r_k^2= (x-x_k)^2 +(y-y_k)^2  $. The vortices are located at  
$x_k=  \pi(3/4,  5/4, 5/4)$ and
$y_k =[ 1 ,  1,  1+1/(2\sqrt 2 )]\pi  $.  Further $\beta = 1/\pi$ and $\alpha = \pi $. 
From this the stream function is $\Delta \Psi = \omega$ and finally $(u,v) = (-\Psi_y, \Psi_x) $. 
The Poisson-equation for the stream-function is only solvable if integral condition $\sum \omega = \sum  ( -u_y + v_x ) = 0 $ is fulfilled. 
This determines $\omega_0\approx -0.038$. The time integration is the Gauss second order scheme, i.e. the implicit midpoint rule, with $\Delta t =2.5 \cdot 10^{-3}$.  

The level lines in figure \ref{fig:threeVortex} are chosen in the same manner as in \cite{KevlahanFarge1997}. Fine structures emerge after short time in the vortex merging process.   
The comparison with the cited work is challenging, as it is a comparison with a spectral scheme. However, we find very good agreement with figures presented in that work.  		 

\section{Conclusion}

A novel approach to avoid the odd-even decoupling in the simulation of incompressible flows is presented. It builds on the combination of asymmetric 
derivatives which are found to be momentum and energy conserving if combined appropriately. High order discretization order in space and time can be utilized.
The scheme works on Cartesian and transformed, structured grids.    
Numerical simulations of two dimensional, periodic test configurations are presented, where 
fifth order in space and fourth order in space was used. Energy and momentum conservation was numerically verified. 

%



\end{document}